\documentstyle[aps,prb,multicol,graphicx]{revtex}

\draft

\begin{document}
\title{Superconductivity in the Kondo lattice: a mean-field approach}

\author{M. A. Gusm\~{a}o$^{a}$ and A. A. Aligia$^{b}$}
\address{$^{a}$Instituto de F{\'\i}sica, Universidade Federal do Rio Grande
do Sul, Caixa Postal 15051, 91501-970 Porto Alegre, RS, Brazil
\\ $^{b}$Comisi\'{o}n Nacional de Energ{\'{\i }}a
At\'{o}mica, Centro At\'{o}mico Bariloche and Instituto Balseiro, 
8400 S.C. de Bariloche, Argentina} 

\date{Received \today }
\maketitle

\begin{abstract}
We calculate the superconducting critical temperature $T_{c}$ and the
Kondo temperature $T_{\rm K}$ of the Kondo lattice, decoupling the
Kondo exchange interaction in the unrestricted Hartree-Fock (HF)
Bardeen-Cooper-Schrieffer (BCS) approximation. We obtain that both
$T_{\rm K}$ and $T_{c}$ have an exponential dependence in the Kondo
coupling $J_{\rm K}$. For optimum doping and realistic parameters,
both temperatures fall in the experimentally observed range.
\end{abstract}


\begin{multicols}{2}

\section{Introduction}

One of the main characteristics of heavy-fermion systems\cite{hewson}
is the competition between Kondo effect and magnetic long-range
order,\cite{expe,doni} both stemming from the local antiferromagnetic
(Kondo) coupling between localized moments and conduction-electron
spins. The non-magnetic compounds, like CeAl$_3$, present a
low-temperature state resembling a Fermi liquid with enhanced carrier
mass, the so called heavy-fermion regime. In addition to that,
superconductivity is observed below $T_{c}\sim 1$K in some
systems,\cite{stewart,trova} the best known being CeCu$_2$Si$_2$,
UBe$_{13}$, and UPt$_3$. Although the nature of superconductivity in
these systems is still unclear, it has been suggested that the Kondo
interaction itself might provide a pairing mechanism.\cite{kaga}
Indeed, there is experimental evidence that heavy electrons are
involved in the superconducting state, as a high value is
observed for the specific-heat jump at the superconducting transition.

Lately, the competition between Kondo effect and magnetism has been
studied in a mean-field approximation based on a simple decoupling of
the Kondo exchange term,\cite{iglesias,MAG} but superconductivity has not been
studied within a similar approach. If fluctuations are not included,
this mean-field approximation is equivalent to considering only part
of the terms of an unrestricted Hartree-Fock (HF) decoupling of the
exchange interaction.  In this work, we include all relevant terms in
this HF decoupling. Among them, a singlet pairing
coupling appears, which we treat in the Bardeen-Cooper-Schrieffer (BCS)
approximation, leading to superconductivity in the model. Although
there are experimental grounds to the notion that superconductivity in
heavy-fermion systems presents non-trivial symmetry
properties,\cite{stewart} and can even coexist with magnetic ordering,
it is instructive to explore, in a simple theoretical approach, the
existence of a superconducting solution in which the pairing mechanism
is the same local Kondo interaction that is also responsible for
magnetic ordering and compensation of local moments.

\section{Model and mean-field approximation}

We consider the standard Kondo-lattice model used to describe
heavy-fermion systems, writing the Hamiltonian in the form
\begin{equation}
H=-\,t \sum_{\langle ij\rangle \sigma } (c_{i\sigma }^{\dagger}
c_{j\sigma} + \text{H.c.})  + J_{\rm K}\sum_{i}{\bf S}_{i}\cdot {\bf
s}_{i}, \label{h}
\end{equation}
where $c_{i\sigma }^{\dagger }$ creates a conduction electron with
spin $\sigma $ at site $i$, and ${\bf S}_{i}$ (${\bf s}_{i}$) is the
localized (conduction) spin operator at site $i$.

The HF-BCS approximation has been described before, \cite{swave} and
has been used, for example, to study $d$-wave superconductivity in a
generalized Hubbard model,\cite{dwave} with results which are in
qualitative agreement with exact diagonalization of finite
systems.\cite{num} We use a fermionic representation of the localized
spins, assigning them to $f$ electrons, whose occupation number is
restricted to one per site. Then, neglecting expectation values which
do not conserve spin (e.g, triplet pairing), the unrestricted HF
decoupling of the interaction term reads 
\end{multicols}
\vspace*{-12pt}
\noindent\rule{8.7cm}{0.7pt}\rule{0.7pt}{6pt}
\begin{eqnarray}
{\bf S}_{i}\cdot {\bf s}_{i} &=&
\frac{1}{4}(f_{i\uparrow }^{\dagger }f_{i\uparrow
}^{{}}-f_{i\downarrow }^{\dagger }f_{i\downarrow }^{{}})(c_{i\uparrow
}^{\dagger }c_{i\uparrow }^{{}}-c_{i\downarrow }^{\dagger }c_{i\downarrow
}^{{}}) +
\frac{1}{2} (f_{i\uparrow}^{\dagger} f_{i\downarrow}^{} 
c_{i\downarrow}^{\dagger} c_{i\uparrow }^{} + \text{H.c.})  
\nonumber \\
&\simeq &
\langle S_{i}^{z} \rangle s_{i}^{z} + S_{i}^{z} 
\langle s_{i}^{z} \rangle - \langle S_{i}^{z} \rangle 
\langle s_{i}^{z} \rangle  
+ \frac{1}{4} \sum_{\sigma} \left( 
\langle f_{i\sigma}^{\dagger} c_{i\sigma}^{} \rangle
f_{i\sigma}^{} c_{i\sigma}^{\dagger} 
+ f_{i\sigma}^{\dagger} c_{i\sigma}^{} 
\langle f_{i\sigma} c_{i\sigma}^{\dagger} \rangle 
- \langle f_{i\sigma}^{\dagger} c_{i\sigma}^{} \rangle 
\langle f_{i\sigma}^{} c_{i\sigma}^{\dagger} \rangle 
\right.   \nonumber \\ &&  \left. \mbox{} 
+ 2 \langle f_{i\sigma}^{\dagger} c_{i\sigma}^{} \rangle 
f_{i\bar\sigma}^{} c_{i\bar\sigma}^{\dagger} 
+ 2 f_{i\sigma}^{\dagger} c_{i\sigma}^{} 
\langle f_{i\bar\sigma}^{} c_{i\bar\sigma}^{\dagger} \rangle
 - 2 \langle f_{i\sigma}^{\dagger} c_{i\sigma}^{} \rangle 
\langle f_{i\bar\sigma}^{} c_{i\bar\sigma}^{\dagger} \rangle
\right. \nonumber \\ && \left. \mbox{}
+ 2 \langle f_{i\sigma}^{\dagger} c_{i\bar\sigma}^{\dagger} \rangle 
c_{i\sigma}^{} f_{i\bar\sigma}^{}
+ 2 f_{i\sigma}^{\dagger} c_{i\bar\sigma}^{\dagger} 
\langle c_{i\sigma}^{} f_{i\bar\sigma}^{} \rangle 
- 2 \langle f_{i\sigma}^{\dagger} c_{i\bar\sigma}^{\dagger} \rangle 
\langle c_{i\sigma}^{} f_{i\bar\sigma}^{} \rangle 
\right. \nonumber \\ && \left. \mbox{}
+ \langle f_{i\sigma}^{\dagger} c_{i\bar\sigma}^{\dagger} \rangle 
f_{i\sigma}^{} c_{i\bar\sigma}^{}
+ f_{i\sigma}^{\dagger} c_{i\bar\sigma}^{\dagger}
\langle f_{i\sigma}^{} c_{i\bar\sigma}^{} \rangle 
- \langle f_{i\sigma}^{\dagger} c_{i\bar\sigma}^{\dagger} \rangle 
\langle f_{i\sigma}^{} c_{i\bar\sigma}^{} \rangle \right) ,
\label{hf}
\end{eqnarray}
\begin{multicols}{2} \noindent
where $\bar\sigma \equiv -\sigma$. The constraint of one localized
($f$) particle per site is taken only on average, i.e.,
\begin{equation}
\frac{1}{N} \sum_{i\sigma }\langle f_{i\sigma }^{\dagger }f_{i\sigma
}\rangle = 1,
\label{co}
\end{equation}
$N$ being the number of sites, and is imposed by adding to the
Hamiltonian the term
\begin{equation}
H_{c}=E_{f}\sum_{i\sigma }f_{i\sigma }^{\dagger }f_{i\sigma },  \label{hco}
\end{equation}
where $E_{f}$ is a Lagrange multiplier. This procedure reduces the
Hamiltonian to an effective one-body problem.

In the following we consider only the non-magnetic homogeneous
solution with singlet superconductivity. The order parameters are
$\lambda =\langle c_{i\sigma }^{\dagger }f_{i\sigma }\rangle $,
characterizing the singlet Kondo state,\cite{MAG} and $\eta =\langle
f_{i\uparrow }^{\dagger }c_{i\downarrow }^{\dagger }\rangle $, which
is the usual superconducting order parameter. Both can be made real
through a gauge transformation. 

In absence of superconductivity ($\eta =0$), the problem is easily
solved. The electronic structure of $H+H_{c}$ is composed of two
hybrid bands with energies
\begin{equation}
E_{\rm K}^{a(b)}=\frac{\varepsilon _{\rm K}^{}+E_{f}}{2}-(+)\sqrt{\left(
\frac{\varepsilon _{\rm K}^{}-E_{f}}{2}\right) ^{2}+V^{2}}, \label{ek}
\end{equation}
where $\varepsilon _{\rm K}^{}$ is the dispersion relation of the
unperturbed conduction band, and
\begin{equation}
V=\frac{3}{4}J_{\rm K}\lambda .  \label{v}
\end{equation}
Both, $E_{f}$ and $\lambda $ must be determined selfconsistently.  The
latter shows the usual behavior of a mean-field order parameter as a
function of temperature, going to zero at a critical temperature which
we identify as the (mean-field) Kondo temperature $T_{\rm K}$.  In the
limit $\lambda \rightarrow 0^{+}$, the self consistent conditions lead
to a simple equation for $T_{\rm K}$:
\begin{equation}
1=\frac{3}{8N}J_{\rm K}\sum_{\rm K}\frac{\tanh (\frac{ \varepsilon
_{\rm K}^{} - \mu
}{T_{\rm K}})}{ \varepsilon _{\rm K}^{} -\mu }. \label{tk}
\end{equation}
The chemical potential $\mu $ in this limit is determined by the
number of conduction electrons of the unperturbed band,
\begin{equation}
n_{c}= \sum_{\sigma }\langle c_{i\sigma }^{\dagger }c_{i\sigma
}\rangle = 2\int \rho _{0}^{}(\varepsilon ) f(\varepsilon ) d\varepsilon ,  
\label{nc}
\end{equation}
with $f(\varepsilon )$ standing for the Fermi function, and $\rho
_{0}^{}(\varepsilon )$ being the unperturbed density of states (DOS) per
spin. In the following we take, for simplicity, a parabolic DOS,
$\rho _{0}^{}(\varepsilon )=3(1-\varepsilon ^{2})/4$, choosing the half band
width of the unperturbed band as the unit of energy.

Since $H$ is electron-hole symmetric, we can assume, without loss of
generality, $n_{c}\leq 1$, so that the Fermi level falls in the lower
hybrid band. The pairing terms in $H$ can be written in terms of the
hybrid operators describing the quasi-particles in the lower band, and
we can safely neglect the upper one, as the superconducting critical
temperature $T_c$ turns out to be much lower than $T_{\rm K}$. After
some algebra, we arrive at the following four equations, which
determine $E_{f}$, $\lambda $, $\mu $, and the superconducting
critical temperature $T_{c}$:
\begin{eqnarray}
n_{f} &=&\sum_{\sigma }\langle f_{i\sigma }^{\dagger }f_{i\sigma
}\rangle = \frac{2}{N} \sum_{\rm K} [Y_{\rm K}^{2} f(E_{\rm K}^{a}) +
X_{\rm K}^{2} f(E_{\rm K}^{b})], \nonumber \\ \lambda
&=&\frac{V}{2N}\sum_{\rm K}\frac{1}{r_{\rm K}}[f(E_{\rm
K}^{a})-f(E_{\rm K}^{b})], \nonumber \\ n_{f} &+& n_{c}
= \frac{2}{N}\sum_{\rm K}[f(E_{\rm K}^{a})+f(E_{\rm K}^{b})],
\nonumber \\ 1 &=& \frac{3J_{\rm K}V^{2}}{16N} \sum_{\rm K}
\frac{1}{r_{\rm K}^{2} (E_{\rm K}^{a}-\mu )} \tanh \left( \frac{E_{\rm
K}^{a}-\mu } {T_{c}}\right) ,
\label{ts}
\end{eqnarray}
where $X_{\rm K}^{2}=1/2+(E_{f}-\varepsilon _{\rm K})^{2}/(4r_{\rm K})$,
$Y_{\rm K}^{2}=1-X_{\rm K}^{2}$, and $r_{\rm K}=[(E_{f}- \varepsilon_{\rm
K})^{2}/ 4 + V^{2}]^{1/2}$.

\section{Results}
In Fig.\ 1 we present $T_{{\rm K}}$ and the superconducting critical
temperature $T_{c}$ as a function of conduction-band filling $n_{c}$
for a typical value of $J_{{\rm K}}$. The Kondo temperature $T_{{\rm
K}}$ increases with $n_{c}$ and has its maximum at $n_{c}=1$ (as
explained above, $ T_{{\rm K}}(n_{c})=T_{{\rm K}}(2-n_{c})$). Instead,
$T_{c}$ should vanish at $n_{c}=1$ because this case corresponds to a
Kondo insulator, with the Fermi level falling in a gap whose magnitude
is larger than the effective pairing interaction. In other words,
while the effective pairing interaction and the density of states
increase as $n_{c}$ tend to one, the effective number of carriers
around the Fermi level tends to zero. As a consequence of this
competition, $T_{c}$ shows a maximum around $n_{c}\sim 0.65$--0.7,
whose position is not strongly dependent on the coupling, as shown in
Fig.\ 2. Assuming that the magnitude of the unperturbed half band
width is of the order of 1$ \; $eV, then, for $J_{{\rm K}}\sim
0.3\;$eV we obtain $T_{c}\sim 1\;$K at optimum doping (see Fig.\ 1).

\begin{figure}
\includegraphics[width=8cm]{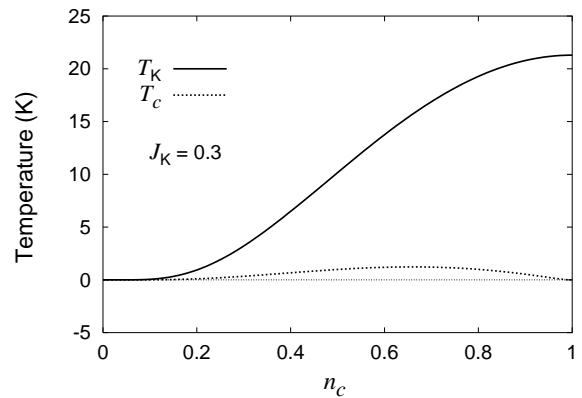}
\caption{Kondo temperature $T_{\rm K}$ and superconducting critical
temperature $T_{c}$ as a function of doping for $J_{\rm K}=0.3$.}
\end{figure}

\end{multicols}

\begin{multicols}{2}

\begin{figure}
\includegraphics[width=8cm]{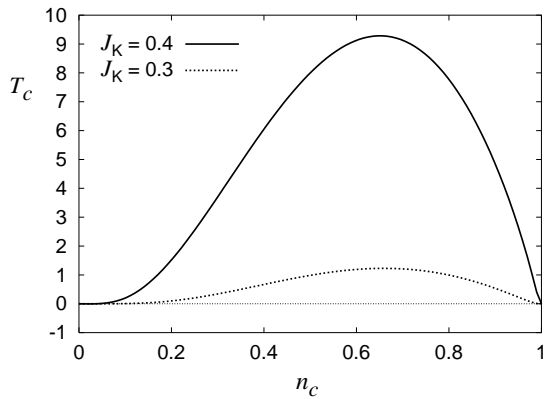}
\caption{Superconducting critical temperature $T_c$ as a function of
 band filling for two different values of the Kondo coupling.}
\end{figure}

\begin{figure}
\includegraphics[width=8cm]{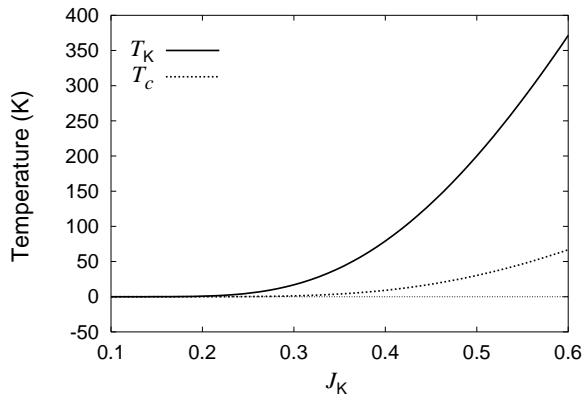}
\caption{Kondo temperature $T_{\rm K}$ and superconducting critical
temperature $T_c$ as functions of $J_{\rm K}$ for $n_{c}=0.7$.}
\end{figure}

\end{multicols}

\begin{multicols}{2}

Figure 3 shows $T_{\rm K}$ and $T_{c}$ as functions of $J_{\rm K}$
for $n_{c}=0.7$ (near optimum doping). The dependence of both
temperatures with $J_{\rm K}$ can be fitted very well by exponential
functions:
\begin{eqnarray}
T_{\rm K} &=& A \exp (-\alpha/J_{\rm K}),  \nonumber \\
T_{c} &=& B \exp (-\beta/J_{\rm K}),  \label{exp}
\end{eqnarray}
as clearly evidenced by the logarithmic plots of Fig.\ 4. 
The meaning of the coefficients is not obvious. The parameter values
used in Figs.\ 2 and 3 yield $A=0.7289$, $\alpha=1.876$, $B=0.3250$,
and $\beta=2.425$.  For a single Kondo impurity, in the limit of
infinite band width, it is known that $T_{\rm K}^{\rm imp}=(1/\rho
_{0}^{})\exp [-1/(\rho _{0}^{}J_{\rm K})]$, \cite{and} where $\rho_{0}^{}$ is
the magnitude of the unperturbed density of states, assumed
constant. The natural extension to the case of a finite band width
would be to write the single-impurity Kondo temperature as the first
of equations (\ref{exp}) with $\alpha^{\rm imp}=1/\rho_0^{} (\varepsilon_{\rm
F}^{})$, which would have the value 1.3905 for the example considered.
However, in the present case the coefficients are not only affected by
the finite band width but also by the mean-field approximation and the
fact that we are dealing with a Kondo lattice instead of a single
impurity. The dependence of $T_{c}$ with $J_{\rm K}$ is even more
difficult to predict. From usual BCS theory one might expect an
exponential dependence on the product of the interacting density of
states and the effective pairing interaction, if the latter is
small. This in turn, is proportional to $J_{\rm K}$ and $\lambda ^{2}$
(see Eq.(\ref{ts})), with a prefactor that, being generally of the
order of magnitude of the band width, should vanish as
$n_{c}\rightarrow 1$.

\section{Discussion}

We have shown that the Kondo-lattice model, treated within the HF BCS
approximation, can lead to a superconducting phase within the Kondo
(heavy-fermion) regime, with the Kondo temperature $T_{\rm K}$ and
superconducting critical temperature $T_{c}$ of the order of magnitude
observed in experiments. While both temperatures depend exponentially
on the exchange interaction, their doping dependences are different,
and $T_{c}$ vanishes for a half-filled band. If one wants to make
quantitative comparisons with experimental results, an appropriate
choice of parameters that gives a good agreement for $T_c$ will yield
a slightly overestimated $T_{\rm K}$. This is typical of a mean-field
solution, and one expects the true Kondo temperature to be reduced by
fluctuation effects. Actually, we would like to remark that $T_{\rm
K}$ signals a second-order phase transition in the mean-field
approach, while only broad features characteristic of a crossover
behavior are observed in experiments.

It is interesting to notice that other approximations can be obtained
if one explicitly uses the constraint $n_{i}^f=1$ to rewrite parts of
the Hamiltonian before performing the HF decoupling. For example
$n_{i}^f n_{i}^c = n_{i}^c$ ceases to be an exact equality once the
first term is decoupled. The use of different approximations affect
the strength of the different decoupled terms, and therefore affect
the relative stability of the different phases. However, assuming that
magnetic ordering does not occur, the qualitative behavior of $T_{\rm
K}$ and $T_{c}$ should be the same as in our approach.

\begin{figure}
\includegraphics[width=8cm]{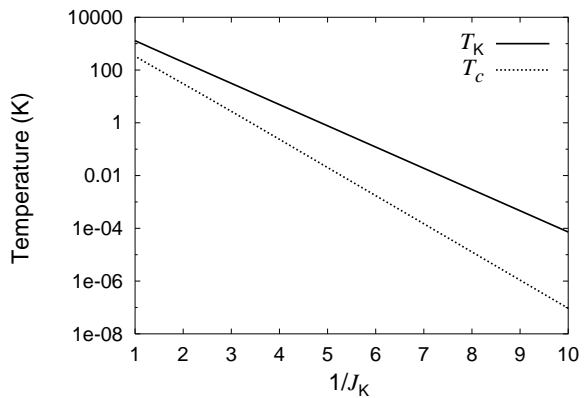}
\caption{Logarithmic plots of $T_{\rm K}$ and $T_{c}$ as functions of
$1/J_{\rm K}$ for $n_{c}=0.7$, showing the straight lines that fit
Eqs.~(\protect\ref{exp}).}
\end{figure}

We have not analyzed here the stability of our solution with respect
to magnetic ordering. However, previous results\cite{mag2} have shown
that within a mean-field approach one needs an explicit intersite
exchange in order to obtain magnetic long-range ordering. Even in that
case, the non-magnetic solution should be stable for moderate or large
Kondo interaction $J_{\rm K}$. On the other hand, the possibility of
obtaining a wave-vector dependent superconducting gap, as suggested by
experimental results,\cite{stewart} has more to do with the model
itself, as one should consider a non-local Kondo interaction. The
present approach can be easily extended to such a case.

\section*{Acknowledgments}

This work benefitted from the grants FINEP-PRONEX 41.96.0907.00
(Brazil), PICT 03-00121-02153 of ANPCyT (Argentina), and PIP 4952/96
of CONICET (Argentina), and the Brazil-Argentina cooperation agreement
CAPES-SECyT 18/00. One of us (M. A. G.) acknowledges partial support
by CNPq (Brazil), and A. A. A.  is partially supported by CONICET.

\end{multicols}

\end{document}